\documentclass[12pt,english]{paper}
\usepackage{amsthm,amsmath,amssymb}
\numberwithin{equation}{section}
\usepackage[T1]{fontenc}
\usepackage[latin1]{inputenc}
\usepackage{graphicx}
\usepackage{cite}
\usepackage{slashed}
\makeatletter

\newcommand{\bibstyle@aas}{\bibpunct{(}{)}{;}{a}{}{,}}

\makeatletter

\makeatletter

\usepackage{geometry}

\makeatletter

\usepackage{epsfig}

\makeatother

\makeatother

\makeatother

\usepackage{babel}
\makeatother
\begin{document}

\title{Measurement of Downward-going Milli-charged particles beyond GZK cutoff at the Pierre Auger Observatory}

\author{Ye Xu}

\maketitle

\begin{flushleft}
School of Electrical and Information Engineering, Fujian University of Technology, Fuzhou 350118, China
\par
e-mail address: xuy@fjut.edu.cn
\end{flushleft}

\begin{abstract}
It is assumed that superheavy dark matter particles (SHDM, $\phi$) with $\mathcal{O}$(ZeV) mass may decay to relativistic milli-charged particles (MCPs, $\chi$) via a channel $\phi\to\chi\bar{\chi}$. The downward-going MCPs passing through the atmosphere can be searched for at the Pierre Auger observatory (Auger). The massless hidden photon model is taken for MCPs to interact with nuclei, so that we evaluated the numbers and fluxes of expected MCPs at Auger assuming 14 years of Auger data. The hybrid data of Auger was fitted with the flux of UHE MCPs exceeding GZK-cutoff energy. Then the corresponding upper limits on $\epsilon^2$ are calculated at 90\% C. L.. These results indicate that MCPs can be searched for with $\epsilon^2\gtrsim 3.73\times10^{-5}$ at Auger, when $m_{\phi}=2\times10^{21}$ eV and $\tau_{\phi}=10^{27}$ s. And a new region of 10$^{10}$ eV < $m_{MCP}$ < 10$^{11.6}$ eV and $\epsilon$ > $2.02\times10^{-2}$ is ruled out in the $m_{MCP}$-$\epsilon$ plane with 14 years of Auger data. These results indicate potential existence of MCPs and SHDM in the Universe.
\end{abstract}


\section{Introduction}
Dark matter (DM) remains a pivotal subject of study in particle physics, cosmology, and astrophysics, with robust observational evidence confirming its existence across multiple scales\cite{bergstrom,BHS}. Cosmological analyses reveal that approximately 84\% of the matter content in the universe is composed of non-baryonic DM \cite{Planck2015}, which forms a diffuse halo structure around galaxies. The local density of this halo is estimated at $\sim$0.3 GeV/cm$^3$,  and its bulk motion relative to the Solar System occurs at velocities of $\sim$230 km/s\cite{JP}. While DM's gravitational imprint is well-established, detection of thermal DM particles remains elusive despite extensive experimental efforts\cite{XENON1T,PANDAX,fermi,antares-icecube-dm-MW,icecubedm-sun,antaresdm-sun,CAST,GlueX,NGC1275,Chooz,dayabayMINOS}.
\par
In this context, Milli-charged particles (MCPs) -- fermions carrying a fractional electric charge $\epsilon e$ ($\epsilon\ll1$, where "e" is the elementary charge)--have emerged as a compelling alternative DM candidate\cite{GH,CY,FLN}. Theoretical frameworks often incorporate a hidden U(1) gauge symmetry to mediate interactions between MCPs and Standard Model (SM) nuclei. This interaction arises via kinetic mixing between the massless hidden photon associated with an unbroken "mirror" U(1)' symmetry and the SM photon \cite{Holdom}. The mixing parameter $\epsilon$ governs both the effective electric charge of MCPs and the coupling strength between the two photon fields. Current constraints on $\epsilon$ are derived from a multifaceted approach: cosmological and astrophysical probes, accelerator-based experiments, precision measurements of ortho-positronium decay and Lamb shifts, and dedicated DM detection campaigns. These complementary investigations impose stringent bounds on the parameter space of MCP models, refining theoretical predictions while motivating advancements in next-generation detection technologies\cite{CM,DHR,DGR,JR,DCB,SLAC,Xenon,LS,OP,Lamb,SUN,earth}.
\par
A viable multicomponent dark matter (DM) framework posits the coexistence of at least two distinct DM species in the Universe. This scenario features:
\par
1.Non-thermal Superheavy Dark Matter (SHDM) $\phi$: Generated in the early Universe through gravitational particle production mechanisms driven by time-dependent spacetime metrics\cite{KC87,CKR98,CKR99,KT,KCR,CKRT,CCKR,KST,CGIT,FKMY,FKM},  these relics exhibit non-relativistic behavior with masses on the order of ZeV ($m_{\phi}\sim$ ZeV). In the model considered here, these particles dominate the current DM density in Universe.
\par
2.Stable Light MCPs $\chi$: Serving as ultrahigh-energy (UHE) decay products of SHDM via only a channel $\phi\to\chi\bar{\chi}$, these fermions possess masses $m_{\chi}\le$ 1 TeV, forming a subdominant component of the present-day DM budget.
\par
Despite the extreme longevity of $\phi$ (far exceeding the cosmic age t$_0$ $\sim$ 10$^{17}$s\cite{AMO,EIP}), its gradual decay generates a trace population of relativistic MCPs with energies up to $\displaystyle\frac{m_\phi}{2}$. In this scenario, it is assumed that MCPs constitute the predominant component of cosmic primary particles exceeding GZK-cutoff energy that successfully propagate to Earth. Consequently, the contamination above GZK-cutoff energy is neglected in this work. Meanwhile, we also assume the GZK cutoff originates from UHE low-Z primary nuclei and their secondaries (here Z $\le$ 8) losing energy through CMB photon interactions.
\par
The Pierre Auger Observatory (Auger), situated near Malarguxe, Mendoza Province, Argentina, is a large-scale detector array spanning 3,000 km$^2$ designed to study extreme energy cosmic rays (EECRs). It employs a hybrid detection system combining:
\par
1.Surface Detector Array (SD): $\sim$ 1600 water-Cherenkov stations measuring secondary particle cascades at ground level.
\par
2.Fluorescence Detector System (FD): Four atmospheric monitoring stations, each equipped with six optical telescopes, to track the longitudinal development of extensive air showers (EAS) through nitrogen fluorescence emissions\cite{Auger2010}.
\par
In this work, the atmosphere itself is utilized as a detection medium for UHE MCPs produced via SHDM decay ($\phi\to\chi\bar{\chi}$). Downward-going UHE MCPs are detectable through their deep inelastic scattering (DIS) interactions with atmospheric nuclei, a process measurable by the SD and FD of Auger. This detection capability will also be discussed here.
\section{Flux of MCPs which reach the Earth}
Since the lifetime for the decay of SHDM to SM particles is strongly constrained ($\tau_\phi \geq$ $\mathcal{O}$($10^{26}$)s) by diffuse gamma and neutrino observations\cite{EIP,MB,RKP,KKK}, $\tau_{\phi}$ is taken to be set to $\tau_{\phi}$ = $10^{27}$ s.
\par
The total flux, $\psi_{\chi}$, is comprised of galactic and extragalatic components: $\psi_{\chi}=\psi_{\chi}^G+\psi_{\chi}^{EG}$, where $\psi_{\chi}^G$ and $\psi_{\chi}^{EG}$ are the galactic and extragalactic MCP fluxes, respectively. The galactic component's differential flux is expressed as\cite{BLS}:
\begin{center}
\begin{equation}
\frac{d\psi_{\chi}^{G}}{dE_{\chi}}=C_G\frac{dN_\chi}{dE_\chi}
\end{equation}
\end{center}
with
\par
\begin{center}
\begin{equation}
C_G=1.7\times10^{-8}\times\frac{10^{26}s}{\tau_{\phi}}\times\frac{1TeV}{m_{\phi}} cm^{-2}s^{-1}sr^{-1}.
\end{equation}
\end{center}
where E$_{\chi}$ (=$m_{\phi}$/2) and N$_{\chi}$ are the energy and number of MCPs, respectively.
\par
The extragalactic component's differential flux is expressed as\cite{BGG,EIP}:
\begin{center}
\begin{equation}
\frac{d\psi_{\chi}^{EG}}{dE_{\chi}}=C_{EG} \int_0^{\infty}dz\frac{1}{H_0\sqrt{\Omega_{\Lambda}+\Omega_m(1+z)^3}}\frac{dN_\chi}{dE_\chi}[(1+z)E_\chi]
\end{equation}
\end{center}
with
\par
\begin{center}
\begin{equation}
C_{EG}=1.4\times10^{-8}\times\frac{10^{26}s}{\tau_{\phi}}\times\frac{1TeV}{m_{\phi}} cm^{-2}s^{-1}sr^{-1}.
\end{equation}
\end{center}
where E$_{\chi}$ $\le$ $m_{\phi}$/2, z represents the red-shift of the source, $\Omega_{\Lambda}=0.685$ is the dark energy density of the Universe,  $\Omega_m=0.315$ is the pressureless matter density of the Universe and $H_0$=67.8 km/s$\cdot$Mpc is the present day Hubble expansion rate from the PLANCK experiment\cite{Planck2015}.
\section{UHE MCP interactions with nuclei}
Here, we adopt the hidden photon framework\cite{Holdom} to characterize MCPs interacting with nuclei through neutral current (NC) processes. This interaction is mediated through a kinetic mixing mechanism between SM photon field and a massless dark sector U(1) gauge boson. The sole phenomenologically viable interaction compatible with SM gauge symmetries arises through a portal mechanism in the dimension-4 operator: $\displaystyle\frac{\epsilon}{2}F_{\mu\nu}F^{\prime\mu\nu}$ where $F_{\mu\nu}$ denotes the SM electromagnetic field tensor and $F^{\prime\mu\nu}$ represents the field strength tensor of the hidden sector photon. Its interaction Lagrangian can be expressed as follows:
\begin{center}
\begin{equation}
\mathcal{L} =\sum_qe_q\bar{q}\gamma^{\mu}qA_{\mu} -\frac{1}{4}F^{\prime}_{\mu\nu}F^{\prime\mu\nu}+\bar{\chi}(i\slashed{D}-m_{\chi})\chi-\frac{\epsilon}{2}F_{\mu\nu}F^{\prime\mu\nu}
\end{equation}
\end{center}
where $\chi$ denotes a MCP with $\epsilon e$, the sum runs over quark flavors in the nucleon and $e_q$ is the electric charge of the quark. $A_{\mu}$ is the vector potential of the SM photon.  $m_{\chi}$ is the MCP's mass. $\epsilon$ is the kinetic mixing parameter between the SM and hidden photons. The covariant derivative is
\begin{center}
\begin{equation}
D_{\mu}=\partial_{\mu}-ig_{\chi}A^{\prime}_{\mu}
\end{equation}
\end{center}
where $g_{\chi}$ is the gauge coupling of the U(1)$^{\prime}$ and $A^{\prime}_{\mu}$ is the vector potential of the hidden photon.
\par
The DIS cross section for MCPs on nuclei, computed using the model from Ref.\cite{SUN}, scales as $\epsilon^2$ relative to the neutral-current electromagnetic DIS cross section for electrons on nuclei:
\begin{center}
\begin{equation}
\sigma_{\chi N}\approx\epsilon^2\sigma^{\gamma}_{eN}
\end{equation}
\end{center}
where N denotes a nucleon. $\sigma^{\gamma}_{eN}$ denotes the cross section depending on $\gamma$ exchange between elections and nuclei. The doubly differential electromagnetic DIS cross section for electrons on nuclei is expressed via the structure functions\cite{SUN}
\begin{center}
\begin{equation}
\frac{d^2\sigma^{\gamma}_{eN}}{dxdQ^2}=\frac{2\pi\alpha^2}{xQ^4}Y_+F_2^{\gamma}
\end{equation}
\end{center}
where $Q^2$ denotes the momentum transfer, $\alpha$ denotes the fine-structure constant. $Y_+=1+(1-y)^2$, where the inelasticity parameter $y=\displaystyle\frac{Q^2}{2m_NE_{in}}$. $m_N$ denotes the nucleon mass, $E_{in}$ denotes the incident electron energy (also the incident MCP energy). The structure function $F_2^{\gamma}$ can be expressed in terms of the quark and anti-quark parton distribution functions.
\par
The electromagnetic DIS cross sections for electrons on nuclei are tabulated in table II of Ref.\cite{BDH}. These results can be parametrized analytically  at energies above 1 EeV:
\begin{center}
\begin{equation}
\begin{split}
\sigma^{\gamma}_{eN}\approx&1.145\times10^{-30}+2.938\times10^{-31}\rm ln[E_{in}(GeV)]-2.082\times10^{-32}\rm ln^2[E_{in}(GeV)]\\
&+1.205\times10^{-33}\rm ln^3[E_{in}(GeV)]\\
\end{split}
\end{equation}
\end{center}
The MCP and neutrino interaction lengths can be obtained by
\begin{center}
\begin{equation}
L_{\nu,\chi}=\frac{1}{N_A\rho\sigma_{\nu,\chi N}}
\end{equation}
\end{center}
where $N_A$ is the Avogadro constant, and $\rho$ is the density of matter, which MCPs and neutrinos interact with.
\section{Evaluation of the numbers of expected MCPs and neutrinos}
UHE MCPs traversing the atmosphere interact with atmospheric nuclei, generating UHE hadrons through nuclear interactions. These hadronic collisions produce secondary particles that subsequently develop into EAS, whose particle components are predominately electrons and muons propagating through the atmosphere. The Auger Pierre Observatory detects these charged particles through its SD while simultaneously measuring fluorescence radiation emitted by shower particles using FD at energies above 1 EeV. The hybrid detection technique, combining SD and FD data, significantly enhances reconstruction accuracy for both energy determination and angular resolution\cite{ICRC2021_E}. Besides, this methodology operates within Earth's atmospheric envelope, which extends to approximately 100 km altitude.
\par
The number of expected MCPs, N$_{det}$, is expressed as
\begin{center}
\begin{equation}
N_{det} = R\times \int_T \int^{E_{max}}_{E_{min}} \int_{\Omega} \int_{S_{eff}} \Phi_{\chi} P(E,D(\theta_z))cos(\theta_z) dS d\Omega dE dt
\end{equation}
\end{center}
where $\Omega$ is the solid angle and $\theta_z$ is the zenith angles between 0$^{\circ}$ and 60$^{\circ}$. R is the duty cycle for FD and taken to be $\sim$13\%\cite{ICRC2021_E}. T is the lifetime of taking data for Auger and taken to be 14 years in the present work. dS=dx$\times$dy is the horizontal surface element. E is the energy of an incoming particle and varies from $E_{min}$ to $E_{max}$. Here $\Phi_\chi=\displaystyle\frac{d\psi_\chi}{dE_{\chi}}$. $S_{eff}=\int_{A_{gen}} \eta dS$ is the effective observational area for Auger. In this analysis, the detection efficiency $\eta$ at Auger is assumed to depend solely on the primary particle energy \cite{ICRC2005}. $A_{gen}$ are the total areas where shower events hit ground level. $S_{eff}$ can be obtained from Ref.\cite{auger2011}:
\begin{center}
\begin{equation}
\begin{aligned}
 &\mathcal{A}=\displaystyle\frac{d\mathcal{E}}{dt} =\int_{\Omega} \int_{A_{gen}} \eta cos(\theta_z) dS d\Omega \\
 &= \int_{A_{gen}} \eta dS \int_{\Omega} cos(\theta_z) d\Omega = S_{eff}\int_{\Omega} cos(\theta_z) d\Omega \\
 &S_{eff}=\displaystyle\frac{\mathcal{A}}{\int_{\Omega} cos(\theta_z) d\Omega}
\end{aligned}
\end{equation}
\end{center}
where $\mathcal{A}$ is the aperture of the Auger detector. $\theta_z$ is the zenith angle at Auger, and varies from $0^{\circ}$ to $60^{\circ}$. $\Omega$ is the solid angle. $\mathcal{E}$ is the hybrid exposure of the Auger detector (see figure 11 in Ref.\cite{auger2011}). The hybrid data consists of events observed on the FD and that triggered by at least one SD detector. $P(E,D(\theta_z)$ can be obtained by
\begin{center}
\begin{equation}
P(E,D(\theta_z))=1-exp(-\displaystyle\frac{D(\theta_z)}{L_{air}})
\end{equation}
\end{center}
where $L_{air}$ is the MCP interaction lengths with the air and D($\theta_z$) is the effective length in the atmosphere above the Auger array and $D(\theta_z) = \displaystyle\frac{H}{cos(\theta_z)}$.
\section{Results}
The energy spectra of UHE cosmic rays are taken from Ref.\cite{ICRC2021_E} and were measured with 14 years of Auger data (see red trangles and black circles in figure 1). As mentioned above, in the present scenario, primary particles exceeding GZK-cutoff energy that successfully reach Earth consist predominantly of downward-going UHE MCPs produced by decay of SHDM with mass above 1 ZeV. Meanwhile, the GZK cutoff also originates from UHE low-Z primary nuclei and secondaries (here Z $\le$ 8) losing energy via CMB photon interactions. In the present paper, the cosmic-ray energy spectrum in the GZK-cutoff region was fitted using the method described in Ref.\cite{icrc2023}. In addition, this work employed the hybrid data for analysis.
\par
As shown in figure 1, the olive solid line represents the fit (called "Best fit") with the contributions including the fluxes of UHE low-Z (Z $\le$ 8) primary nuclei and their secondaries for the EPOS-LHC hadronic interaction model, when $X_S$ = 2.5 (where $X_S$ = $\displaystyle\frac{d_S}{\sqrt{r_HL_{coh}}}$, with the coherent length $L_{coh}$ and the Hubble radius $r_H=c/H_0$) and the steepness of the cutoff $\Delta$ = 3\cite{icrc2023}. The black solid line represents the fit with the flux of UHE MCPs with $\epsilon^2$ = $1.3\times10^{-4}$, $m_{\phi}$ = $2\times10^{21}$ eV and $\tau_{\phi}$ = $10^{27}$ s exceeding GZK-cutoff energy. The magenta dash line represents the total flux, e.g., the sum of Best fit and MCPs cases. This figure suggests that MCPs may be one of the candidate components making up the cosmic-ray spectrum above the GZK-cutoff energy.
\par
Using the Feldman--Cousins approach \cite{FC}, we also derive the limits on $\epsilon^2$ at 90\% C.L under the assumption that no events are observed in the measurement of downward-going MCPs at the Auger observatory. Figure 2 presents these limits for $m_{\phi} = 2\times10^{22}$ eV (red dashed line) and $2\times10^{21}$ eV (blue dotted line). For $m_{\phi} = 2\ \text{EeV}$, corresponding to an MCP energy of 1 EeV, the region $\epsilon^2 > 9.1\times10^{-5}$ (i.e., $\epsilon > 9.54\times10^{-3}$) is excluded for $\tau_{\psi} = 10^{27}$ s.
\par
The numbers of expected MCPs were evaluated at different energies assuming 14 years of Auger data. Figure 3 presents the numbers of expected MCPs in the 10$^{18}$ eV - 10$^{20.5}$ eV range for $\epsilon^2=1.3\times10^{-4}$ (red solid line) obtained from fits to hybrid data that exceed the GZK cutoff energy at the Auger observatory, $9.1\times10^{-4}$ (blue dash line) computed at 90\% C.L. using the Feldman--Cousins approach, and $3.73\times10^{-5}$ (magenta dash-dot line) under detectable constraint for MCPs at the Auger observatory (i.e. requiring at least one MCP event to be measured at Auger), when $m_{\phi}=2\times10^{21}$ eV. Figure 4 also presents the numbers of expected MCPs in the 10$^{18}$ eV - 10$^{20.5}$ eV range corresponding to $\epsilon^2=4.11\times10^{-2}$ (red solid line), $2.877\times10^{-2}$ (blue dash line) and $1.18\times10^{-2}$ (magenta dash-dot line), when $m_{\phi}=2\times10^{22}$ eV.
\par
The upper limit for $\epsilon^2$ at 90\% C.L. have been calculated with $m_{\phi}$ = $2\times10^{21}$ eV and $\tau_{\phi}$ = $10^{27}$ s. The region of $\epsilon>9.54\times10^{-3}$ is ruled out in the $m_{MCP}$-$\epsilon$ plane, when $m_{MCP}$ < 1 TeV. This result is shown in figure 2. To compare to other observations on MCPs, this figure also shows the $\epsilon$ bounds from cosmological and astrophysical observations\cite{CM,DHR,DGR,JR}, accelerator and fixed-target experiments\cite{DCB,SLAC}, experiments for decay of ortho-positronium\cite{OP}, Lamb shift\cite{Lamb}, measurement of MCPs from the sun's core\cite{SUN}, the constraint on N$_{eff}$ at the CMB epoch by Planck\cite{VR} (N$_{eff}<3.33$\cite{planck2018}), measurement of MCPs from the Earth's core\cite{earth}, measurement of upward-going MCPs\cite{up-Auger} and measurement of MCPs with a running electromagnetic coupling
constant\cite{running}. A new region of 10$^{10}$ eV < $m_{MCP}$ < 10$^{11.6}$ eV and $2.02\times10^{-2}<\epsilon<1$ is ruled out in the $m_{MCP}$-$\epsilon$ plane with the 14 year of Auger data, as shown in figure 5.
\section{Discussion and Conclusion}
Since $\Phi_{MCP}\propto\displaystyle\frac{1}{\tau_{\phi}}$, the above results scale with $\tau_{\phi}$. For $\tau_{\phi}$ ranging from $10^{28}$ s to $10^{29}$ s, the numbers of expected MCPs at Auger are reduced by factors of 10 to 100 compared to the $\tau_{\phi}=10^{27}$ s case.
\par
Based on the results described above, the downward-going MCPs with $m_{\phi}$ = $2\times10^{21}$, $2\times10^{22}$ eV can be searched for at Auger, when $\epsilon^2 \gtrsim 3.73\times10^{-5}$, respectively. Besides, these results suggests that MCPs may be one of the candidate components making up the cosmic-ray spectrum above the GZK-cutoff energy. Since these constraints are only given by the assumptions mentioned above, certainly, the experimental collaborations, like the Pierre Auger collaboration, should be encouraged to conduct an unbiased analysis with the data of Auger.
\par
While SHDM decay into MCPs would inject extra energy during recombination-reionization era -- potentially constraining parameters via early universe observations -- the SHDM lifetime ($\tau_{\phi} \gg t_0$) yields negligible relic abundance:
\begin{center}
\begin{equation}
\Omega_{MCPs} h^2=2(1-exp(-\displaystyle\frac{T_{re}}{\tau_{\phi}}))\Omega_{\phi}h^2\approx2\displaystyle\frac{T_{re}}{\tau_{\phi}}\Omega_{\phi}h^2\lesssim 10^{-14}\Omega_{\phi}h^2
\end{equation}
\end{center}
where $T_{re}$ is the age of the Universe at the recombination epoch. This fall orders of magnitude below the Planck constraint $\Omega_{MCPs}h^2<0.001$\cite{DDRT}. Consequently, current cosmological observations cannot constrain the this scenario's parameters.
\par
The JEM-EUSO space-based fluorescence detector, designed to study UHE cosmic rays, features an extensive observational area $\mathcal{O}$(10$^5$ km$^2$)\cite{jem-euso2023}. Our preliminary assessment indicates its MCP detection capability significantly surpasses that of Auger's FD, with an expected event rate approximately 20 times higher. Direct MCP detection at JEM-EUSO requires $\epsilon^2 \gtrsim \mathcal{O}(10^{-5})$, demonstrating its advantage over Auger's FD for such measurements. This potential, however, remains contingent on the commencement of JEM-EUSO data collection.
\section{Acknowledgements}
This work was supported by the National Natural Science Foundation
of China (NSFC) under the contract No. 11235006, the Science Fund of
Fujian University of Technology under the contracts No. GY-Z14061 and GY-Z13114 and the Natural Science Foundation of
Fujian Province in China under the contract No. 2015J01577.
\par

\newpage

\begin{figure}
 \centering
 \includegraphics[width=0.9\textwidth]{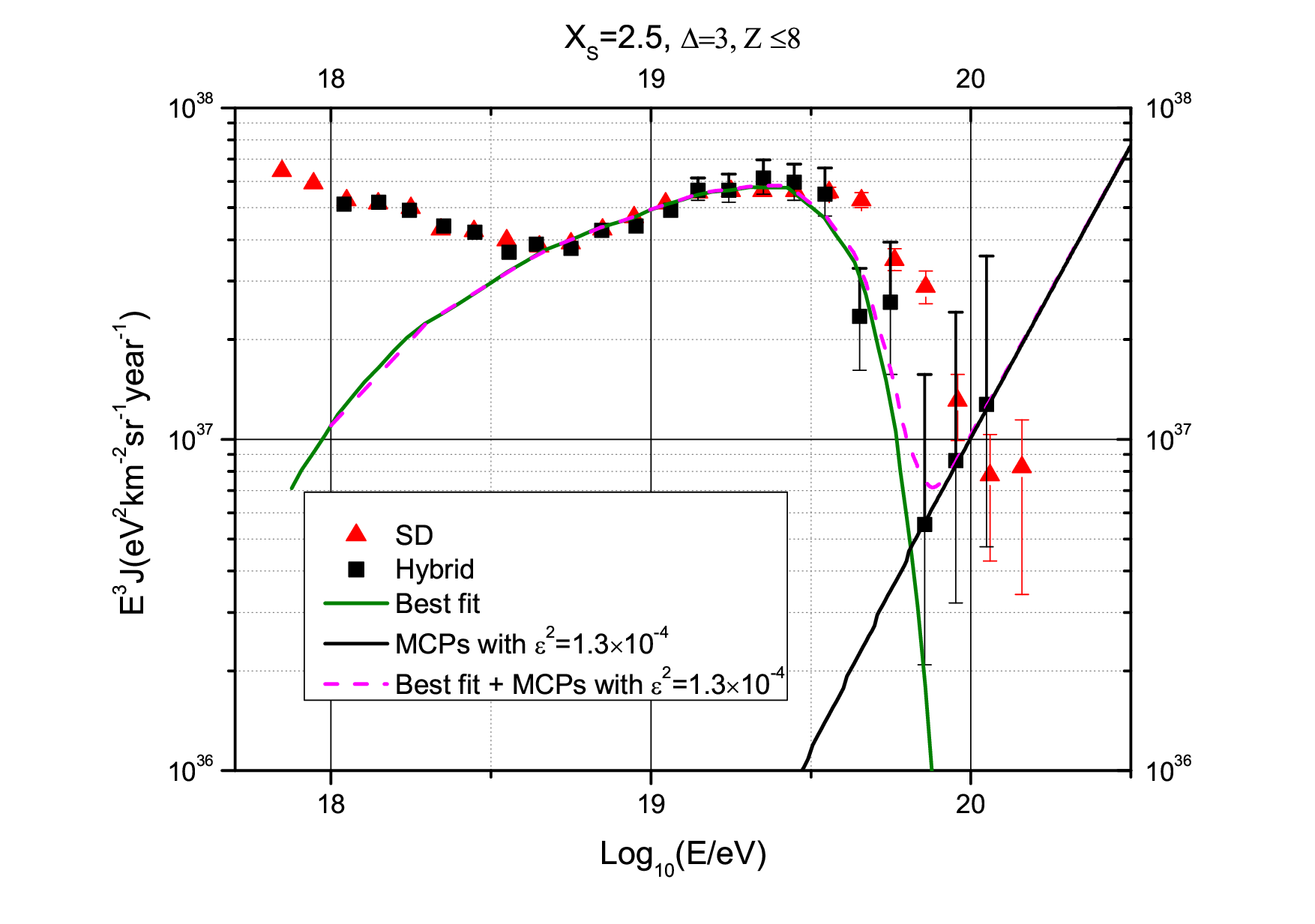}
 \caption{Fits to the fluxes at Earth. The olive solid line represents the fit (called "Best fit") with the contributions including the fluxes of UHE (Z $\le$ 8) primary nuclei and their secondaries, when $X_S$ =2.5 and $\Delta$ = 3. The black solid line represents the fit with the flux of MCPs with $\epsilon^2$ = $1.3\times10^{-4}$, $m_{\phi}$ = $2\times10^{21}$ eV and $\tau_{\phi}$ = $10^{27}$ s exceeding GZK-cutoff energy. The magenta dash line represents the total flux, e.g., the sum of Best fit and MCPs cases.}
 \label{fig:flux}
\end{figure}

\begin{figure}
 \centering
 \includegraphics[width=0.9\textwidth]{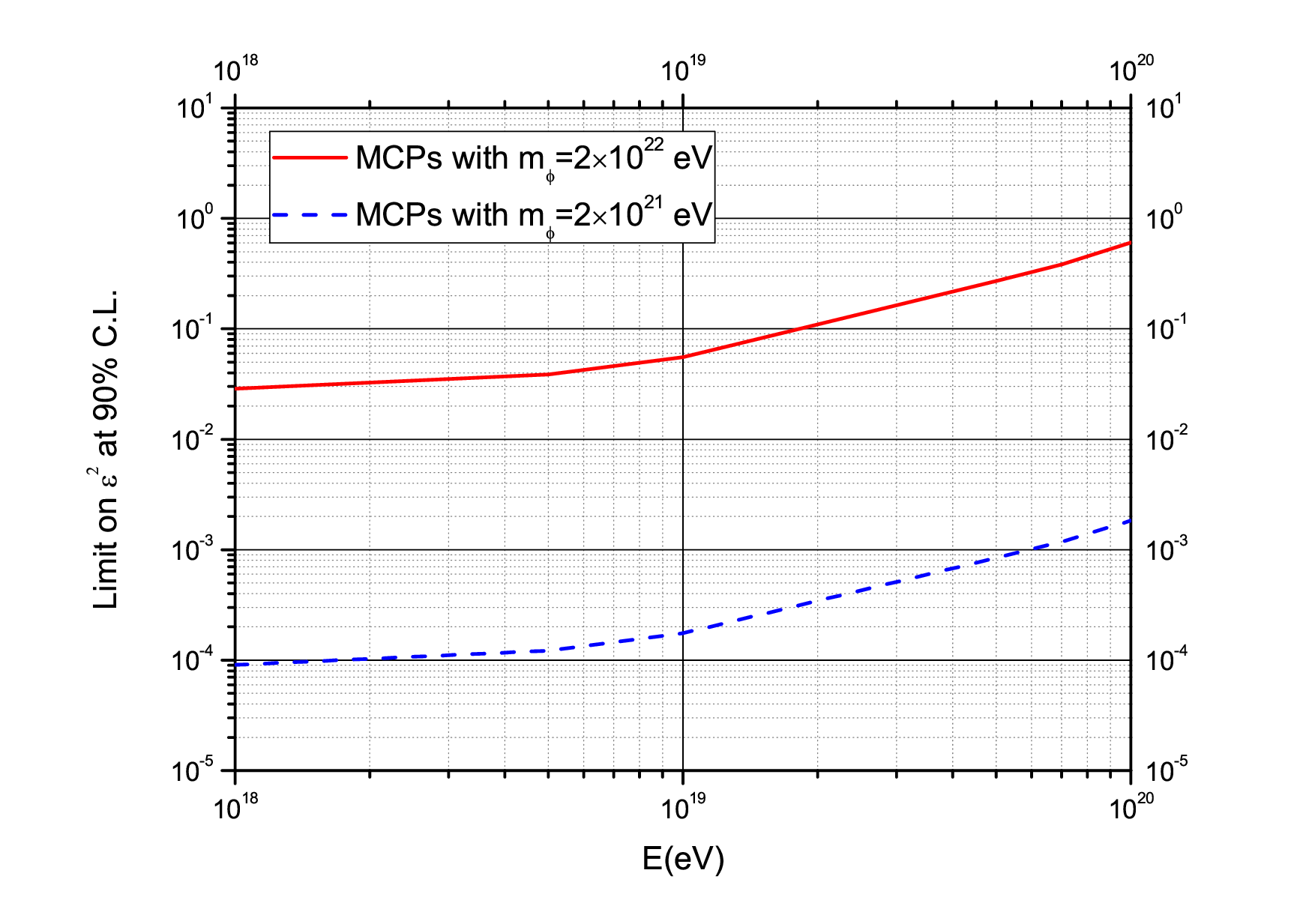}
 \caption{With the different $m_{\phi}$ (= $2\times10^{21}$ eV and $2\times10^{22}$ eV), the limits on $\epsilon^2$ at 90\% C.L. were computed, respectively, assuming no observation at Auger in 14 years.}
 \label{fig:uplimit}
\end{figure}

\begin{figure}
 \centering
 \includegraphics[width=0.9\textwidth]{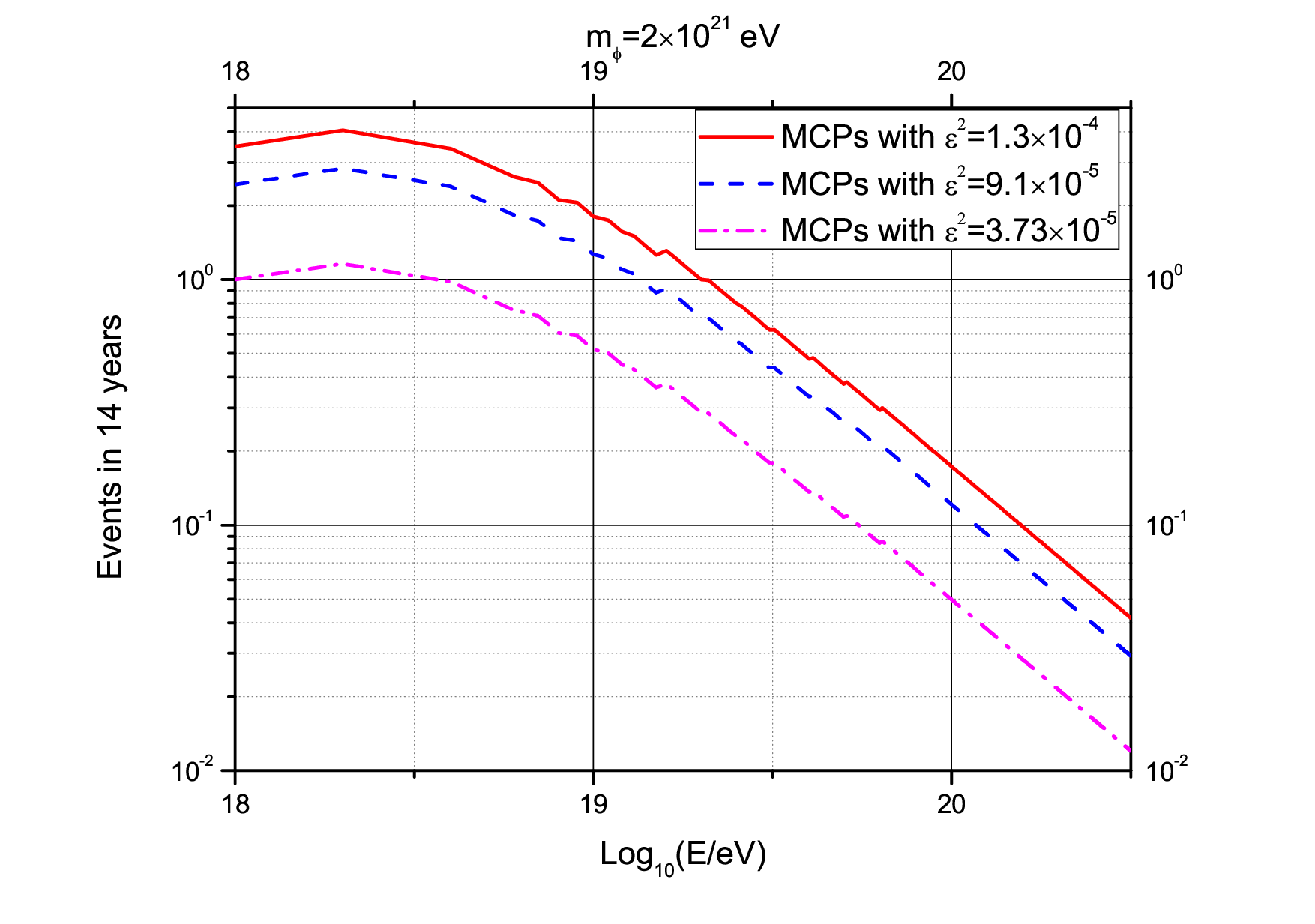}
 \caption{With the different $\epsilon^2$ (= $1.3\times10^{-4}$, $9.1\times10^{-5}$ and $3.73\times10^{-5}$), the numbers of expected MCPs were evaluated assuming 14 years of Auger data, respectively.}
 \label{fig:event_2e21}
\end{figure}

\begin{figure}
 \centering
 \includegraphics[width=0.9\textwidth]{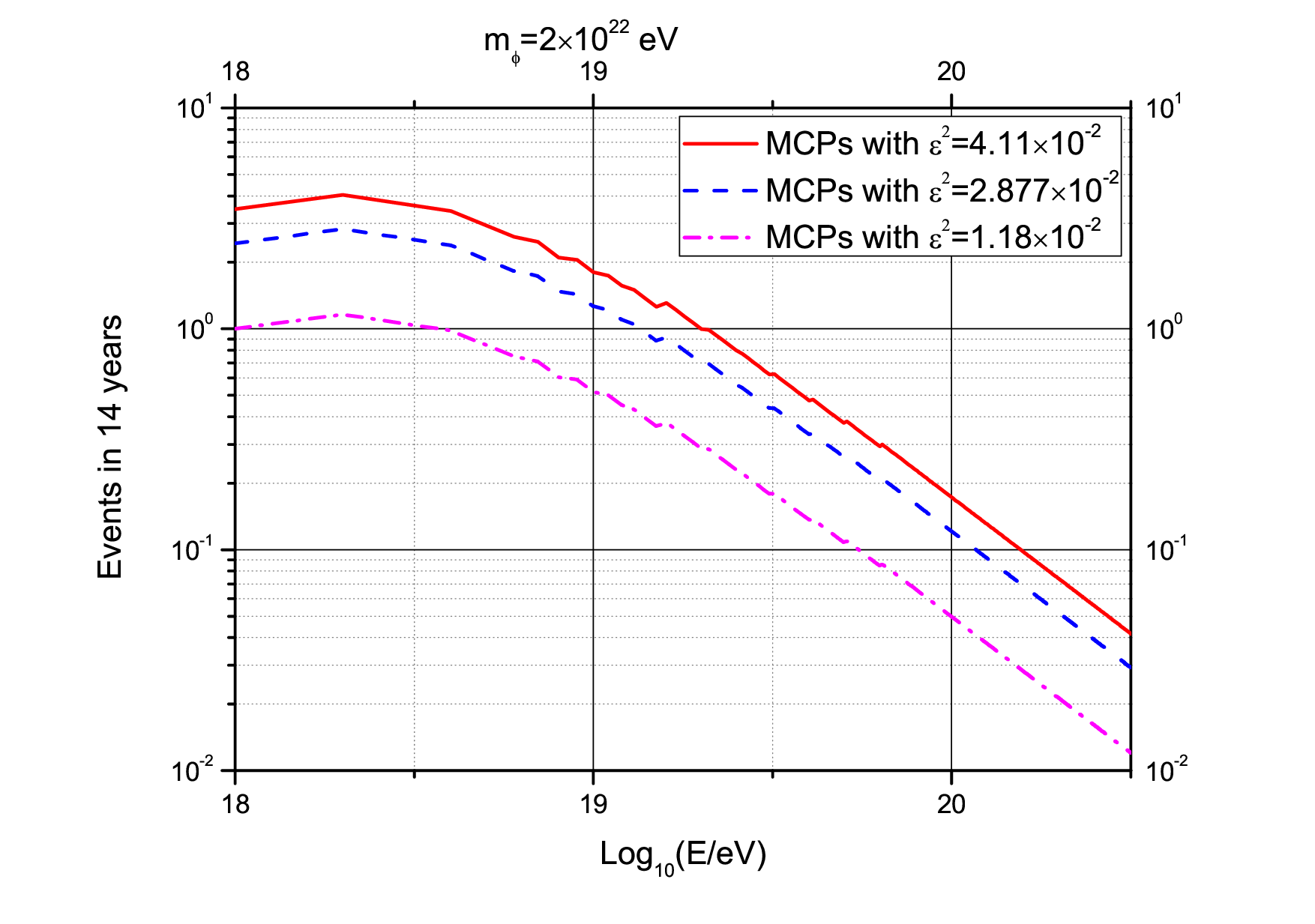}
 \caption{With the different $\epsilon^2$ (= $4.11\times10^{-2}$, $2.877\times10^{-2}$ and $1.18\times10^{-2}$), the numbers of expected MCPs were evaluated assuming 14 years of Auger data, respectively.}
 \label{fig:event_2e22}
\end{figure}

\begin{figure}
 \centering
 \includegraphics[width=0.9\textwidth]{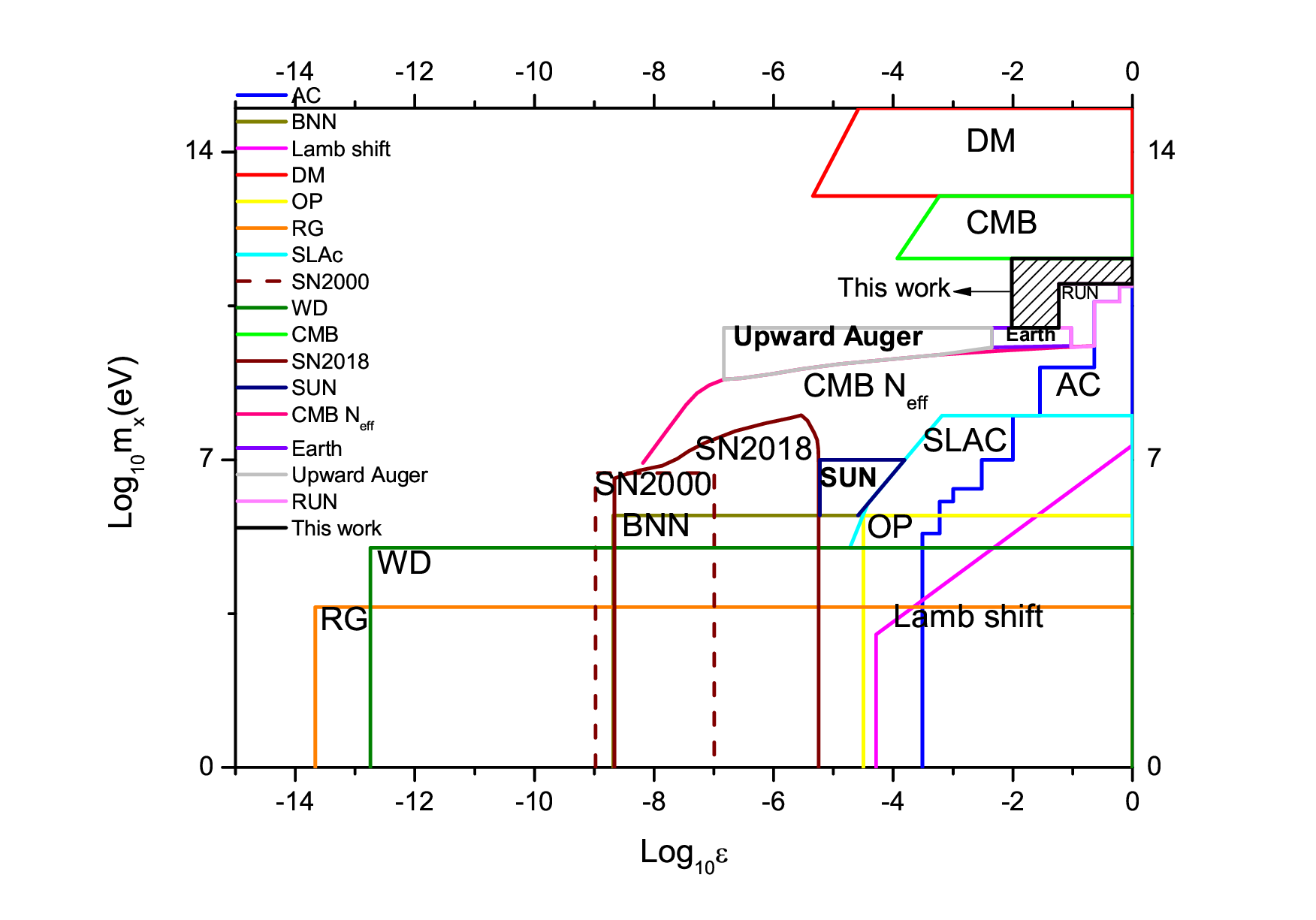}
 \caption{If $m_{\phi}$=$2\times10^{21}$ eV, with $\tau_{\phi}=10^{27}$ s, a new region (shaded region) is ruled out in the $m_{MCP}$ vs. $\epsilon$ plane, when 10$^{9.6}$ < $m_{MCP}$ < 10$^{11.6}$ eV and $\epsilon > 2.02\times10^{-2}$ (this work). Meanwhile, the bounds from plasmon decay in red giants (RG)\cite{DHR}, plasmon decay in white dwarfs (WD)\cite{DHR}, cooling of the Supernova 1987A (SN2000\cite{DHR}, SN2018\cite{CM}), accelerator (AC)\cite{DCB} and fixed-target experiments (SLAC)\cite{SLAC}, the Tokyo search for the invisible decay of ortho-positronium (OP)\cite{OP}, the Lamb shift\cite{Lamb}, big bang nucleosynthesis (BBN)\cite{DHR}, cosmic microwave background (CMB)\cite{DGR}, dark matter searches (DM)\cite{JR}, measurement of MCPs from the Sun's core\cite{SUN}, the constraint on N$_{eff}$ at the CMB epoch by Planck\cite{VR} (N$_{eff}<3.33$\cite{planck2018}), measurement of MCPs from the Earth's core\cite{earth}, measurement of upward-going MCPs\cite{up-Auger} and measurement of MCPs with a running electromagnetic coupling constant\cite{running} are also plotted on this figure.}
 \label{fig:epsilon_bound}
\end{figure}


\begin{thebibliography}{}
\bibitem{bergstrom}L. Bergstrom, Rept. Prog. Phys. 63,793 (2000) arXiv: hep-ph/0002126
\bibitem{BHS}G. Bertone, D. Hooper and J.Silk, Phys. Rep. 405, 279 (2005) arXiv: hep-ph/0404175
\bibitem{Planck2015}P.A.R. Abe, et al., Planck collaboration, A\&A 594, A13 (2015) arXiv:1502.01589
\bibitem{JP}J.D.Lewin, P.F.Smith, Astropart. Phys. 6, 87 (1996)
\bibitem{XENON1T} E.Aprile, et al., XENON1T Collaboration, Phys. Rev. Lett. 119, 181301 (2017)
\bibitem{PANDAX} X.Y.Cui, et al., PandaX-II Collaboration, Phys. Rev. Lett. 119, 181302 (2017)
\bibitem{fermi}M. Ackermann, et al., Fermi-LAT Collaborations, JCAP 09, 008 (2015) arXiv: 1501.05464
\bibitem{antares-icecube-dm-MW} A. Albert, M. Andre, et al., ANTARES and IceCube Collaboration, Phys. Rev. D 102, 08002 (2020) arXiv: 2003.06614
\bibitem{antaresdm-sun}S. Adrian-Matinez, et al., ANTARES Collaboration, Phys. Lett. B 759, 69-74, (2016), arXiv: 1603.02228
\bibitem{icecubedm-sun}M. G. Aartsen, et al., IceCube Collaboration, Euro. Phys. J. C 77, 146 (2017) arXiv: 1612.05949
\bibitem{CAST}V. Anastassopoulos, et al., CAST Collaboration, Nature Physics 13, 584 (2017)
\bibitem{GlueX}S. Adhikari, C.S. Akondi, GlueX Collaboration, Phys. Rev. D 105, 052007 (2022) arXiv: 2109.13439
\bibitem{NGC1275} C.S. Reynolds et al., Astrophys. J. 890, 59 (2020)
\bibitem{Chooz} T. Abrahao, et al., Double Chooz Collaboration, Eur. Phys. J. C 81, 775 (2021) arXiv: 2009.05515
\bibitem{dayabayMINOS}P. Adamson, F.P. An, et al., Daya Bay, MINOS+ Collaboration, Phys. Rev. Lett. 125, 071801 (2020) arXiv: 2002.00301
\bibitem{GH}H. Goldberg and L.J. Hall, Phys. Lett. B 174, 151 (1986)
\bibitem{CY}K.Cheung and T.C. Yuan, JHEP 0703, 120 (2007) arXiv: hep-ph/0701107
\bibitem{FLN}D. Feldman, Z. Liu and P. Nath, Phys. Rev. D 75, 115001 (2007) arXiv: hep-ph/0702123
\bibitem{Holdom}B. Holdom, Phys. Lett. B 166, 196 (1986)
\bibitem{CM}J. H. Chang, R. Essig, and S. D. McDermott, J. High Energy Phys. 09, 051 (2018)
\bibitem{DHR}S. Davidson, S. Hannestad, and G. Raffelt, J. High Energy Phys. 05, 003 (2000)
\bibitem{DGR}S. Dubovsky, D. Gorbunov and G. Rubtsov, JETP Lett. 79 (2004) arXIv: hep-ph/0311189
\bibitem{JR}J. Jaeckel and A. Ringwald, Annu. Rev. Nucl. Part. Sci. 60, 405-437 (2010)
\bibitem{DCB}S. Davidson, B. Campbell and D. Bailey, Phys. Rev. D 43, 2314 (1991)
\bibitem{SLAC}A. Prinz et al., SLAC Collaboration, Phys. Rev. Lett. 81, 1175 (1998) arXiv: hep-ex/9804008
\bibitem{Xenon}R. Essig, T. Volansky, and T.T. Yu, Phys. Rev. D 96, 043017 (2017)
\bibitem{LS}H. Liu and T. R. Slatyer, Phys. Rev. D 98, 023501 (2018)
\bibitem{OP}T. Mitsui et al., Phys. Rev. Lett. 70, 2265 (1993)
\bibitem{Lamb}S.R. Lundeen, F.M. Pipkin, Phys. Rev. Lett. 46, 232 (1981)
\bibitem{SUN} Y. Xu, JHEP 09, 055 (2022) arXiv: 2207.00178
\bibitem{earth}Y. Xu, JHEP (2024) arXiv: 2405.00060
\bibitem{KC87} M.Yu.Khlopov, V.M.Chechetkin, Sov. J. Part. Nucl 18, 267-288 (1987)
\bibitem{CKR98} D. J. H. Chung, E. W.Kolb, and A.Riotto, Phys.Rev.Lett. 81, 4048, (1998) arXiv: hep-ph/9805473
\bibitem{CKR99} D. J. H. Chung, E. W.Kolb, and A.Riotto, Phys.Rev. D59, 023501 (1998) arXiv: hep-ph/9802238
\bibitem{KT}V. Kuzmin and I. Tkachev, JETP Lett. 68, 271¨C275 (1998) arXiv: hep-ph/9802304
\bibitem{KCR} E. W. Kolb, D. J. Chung, and A. Riotto, WIMPzillas!, hep-ph/9810361
\bibitem{CKRT} D. J. H. Chung, E. W. Kolb, A. Riotto, and I. I. Tkachev, Phys. Rev. D 62, 043508 (2000) arXiv: hep-ph/9910437
\bibitem{CCKR}D. J. H. Chung, P. Crotty, E. W. Kolb, and A. Riotto,, Phys. Rev. D 64, 043503 (2001) arXiv: hep-ph/0104100
\bibitem{KST} E. W. Kolb, A. Starobinsky, and I. Tkachev, JCAP 0707, 005 (2007) arXiv: hep-th/0702143
\bibitem{CGIT} L.Covi, M.Grefe, A.Ibarra and D.Tran, JCAP 1004, 017 (2010) arXiv: 0912.3521
\bibitem{FKMY} B.Feldstein, A.Kusenko, S.Matsumoto and T. T.Yanagida Phys. Rev. D 88, 015004 (2013) arXiv: 1303.7320
\bibitem{FKM}M. A. Fedderke, E. W. Kolb, and M. Wyman, Phys. Rev. D 91, 063505 (2015) arXiv:1409.1584

\bibitem{AMO}R. Aloisio, S. Matarrese and A. V. Olinto, JCAP, 1508, 024 (2015), arXiv: 1504.01319
\bibitem{EIP} A.Esmaili, A.Ibarra and O.L. Peres, JCAP, 1211, 034 (2012) arXiv:1205.5281

\bibitem{Auger2010} J. Abraham, et al., The Pierre Auger Collaboration, Phys. Lett. B 685, 239-246 (2010) arXiv: 1002.1975
\bibitem{MB} K.Murase and J.F.Beacom, JCAP 1210, 043 (2012) arXiv:1206.2595
\bibitem{RKP} C.Rott, K.Kohri and S.C.Park, Phys. Rev. D 92, 023529 (2015) arXiv:1408.4575
\bibitem{KKK}M.Kachelriess, O.E.Kalashev and M.Yu.Kuznetsov, Phys. Rev. D 98, 083016 (2018) arXiv: 1805.04500
\bibitem{BLS} Y.Bai, R.Lu and J.Salvado, JHEP, 01, 161 (2016) arXiv:1311.5864
\bibitem{BGG} A.Bhattacharya, R.Gandhi and A.Gupta, JCAP 1503, 027, (2015) arXiv:1407.3280

\bibitem{BDH} M.M. Block, L. Durand and P. Ha, Phys. Rev. D89, 094027 (2014) arXiv: 1404.4530
\bibitem{BHM} Martin M.Block, Phuoc Ha, Douglas W.McKay, Phys. Rev. D 82, 077302 (2010) arXiv: 1008.4555
\bibitem{ICRC2021_E}M. Mastrodicasa on behalf of the Pierre Auger Collaboration, Proceedings of the 37th International Cosmic Ray Conference, Berlin, Germany, 12-23 July 2021, 324
\bibitem{ICRC2005}J.A. Bellido, et al., the Pierre Auger Collaboration, Proceedings of the 29th International Cosmic Ray Conference, Pune, India, 3-10 August 2005, 101-106, arXiv: astro-ph/0507103
\bibitem{auger2011}P. Abreu, et al., the Pierre Auger Collaboration, Astroparticle Physics, 34, 368-381 (2011) arXiv: 1010.6162
\bibitem{icrc2023} J.M. Gonzalez for the Pierre Auger Collaboration, PoS(ICRC2023)288, Proceedings of the 38th International Cosmic Ray Conference, Nagoya, Japan, 26 July - 3 August, 2023, 288

\bibitem{FC}G. J. Feldman, R. D. Cousins, Phys. Rev. D 57, 3873 (1998)
\bibitem{VR} H. Vogel and J. Redondo, JCAP, 1402, 029 (2014) arXiv: 1311.2600
\bibitem{planck2018}N.Aghanim, et al., Planck Cobbaboration, A\&A 641, A6 (2020) arXiv:1807.06209
\bibitem{up-Auger}Y. Xu, J. Lan and W. Gao, Phys. Lett. B 859, 139127, 2024
\bibitem{running}Y. Xu, Measurement of Milli-Charged Particles with a running electromagnetic coupling constant at IceCube, arXiv: 2511.18881
\bibitem{DDRT}A.D. Dolgov, S.L. Dubovsky, G.I. Rubtsov and I.I. Tkachev, Phys. Rev. D 88, 117701 (2013) arXiv: 1310.2376
\bibitem{jem-euso2023} S. Abe, et al., Eur. Phys. J. C 83, 1028 (2023) arXiv: 2311.12656


\end{thebibliography}
\end{document}